\documentclass{article}
\usepackage{spconf,amsmath,graphicx}
\usepackage{booktabs}       
\usepackage{amsfonts}       
\usepackage{multirow}
\usepackage{float}

\title{RobustDistiller: Compressing Universal Speech Representations for Enhanced Environment Robustness}
\name{\begin{tabular}{c}Heitor R. Guimarães$^{1}$, Arthur Pimentel $^{1}$, Anderson R. Avila$^{1}$,  Mehdi Rezagholizadeh$^{2}$, \\ Boxing Chen$^{2}$, Tiago H. Falk$^{1}$\end{tabular}\thanks{Any opinions, findings, and conclusions expressed in this manuscript are those of the authors and do not necessarily reflect the views, official policy or position of Huawei.}}%
    
\address{%
    $^{1}$  INRS-EMT, Université du Québec, Montréal, Canada\\%
    $^{2}$ Huawei Noah’s Ark Lab, Montréal, Canada%
}
\begin{document}
%
\maketitle
\begin{abstract}
Self-supervised speech pre-training enables deep neural network models to capture meaningful and disentangled factors from raw waveform signals. The learned universal speech representations can then be used across numerous downstream tasks. These representations, however, are sensitive to distribution shifts caused by environmental factors, such as noise and/or room reverberation. Their large sizes, in turn, make them unfeasible for edge applications. In this work, we propose a knowledge distillation methodology termed RobustDistiller which compresses universal representations while making them more robust against environmental artifacts via a multi-task learning objective. The proposed layer-wise distillation recipe is evaluated on top of three well-established universal representations, as well as with three downstream tasks. Experimental results show the proposed methodology applied on top of the WavLM Base+ teacher model outperforming all other benchmarks across noise types and levels, as well as reverberation times. Oftentimes, the obtained results with the student model (24M parameters) achieved results inline with those of the teacher model (95M).
\end{abstract}
\section{Introduction}
\label{sec:intro}
Speech representation learning (SRL) and, more specifically, self-supervised learning (SSL) have become ubiquitous methods for speech applications. With such methods, meaningful features from high-dimensional data are learned in a self-supervised manner and later used in different downstream tasks, such as speech recognition or keyword spotting. In fact, as edge speech applications emerge, SRL offers a viable alternative for resource-constrained devices where it is difficult or impossible to run/store different large and complex models, one per application. Today, wav2vec 2.0 \cite{baevski2020wav2vec}, HuBERT \cite{hsu2021hubert}, and WavLM \cite{chen2022wavlm} represent the most widely deployed universal speech representations.

Current universal speech representations, however, have two major limitations when edge applications are envisioned: (i) their performance drops drastically under unseen environmental conditions; and (ii) their large number of parameters make it unfeasible for resource-constrained devices. For instance, the HuBERT representation can range from 95 million to 1 billion parameters \cite{hsu2021hubert}, challenging its deployment on edge devices. To this end, different model compression schemes have been explored with promising results achieved with knowledge distillation, such as the recent DistilHuBERT representation \cite{chang2022distilhubert}. DistilHuBERT starts with the HuBERT Base model (comprised of 12 Transformer layers and 95 million parameters) as the teacher model and relies on a student model with only two Transformer layers, thus compressing the model to roughly one-quarter of the original size. 

On the environmental robustness side, most representations are sensitive to distribution shifts observed within the test data when conditions not seen during training are presented. For example, the work in \cite{huang22b_interspeech} showed that domain adversarial training and data augmentation could improve the robustness of the HuBERT representation to unseen conditions; such a system has been termed ``Robust HuBERT''. To date, however, no system exists that combines both model compression and environment robustness for reliable edge speech applications. This paper aims to fill this gap.


In this work, we propose RobustDistiller, a novel training scheme inspired by the DistilHuBERT recipe that not only compresses the model, but also results in robustness against noise perturbations that are typical of in-the-wild speech applications. In particular, we show that data augmentation via an online contamination process and a multi-task learning denoising step are essential to achieve such benefits. To assess the benefits of the proposed RobustDistiller paradigm, we distill wav2vec 2.0, HuBERT, and WavLM, and show their improved robustness to unseen additive and convolutive noise conditions. When compared to traditional signal-based speech enhancement, the proposed methodology provides substantial gains.

\section{Proposed RobustDistiller Recipe}
\label{sec:pagestyle}

\begin{figure}
        \centering
        \includegraphics[width=\linewidth]{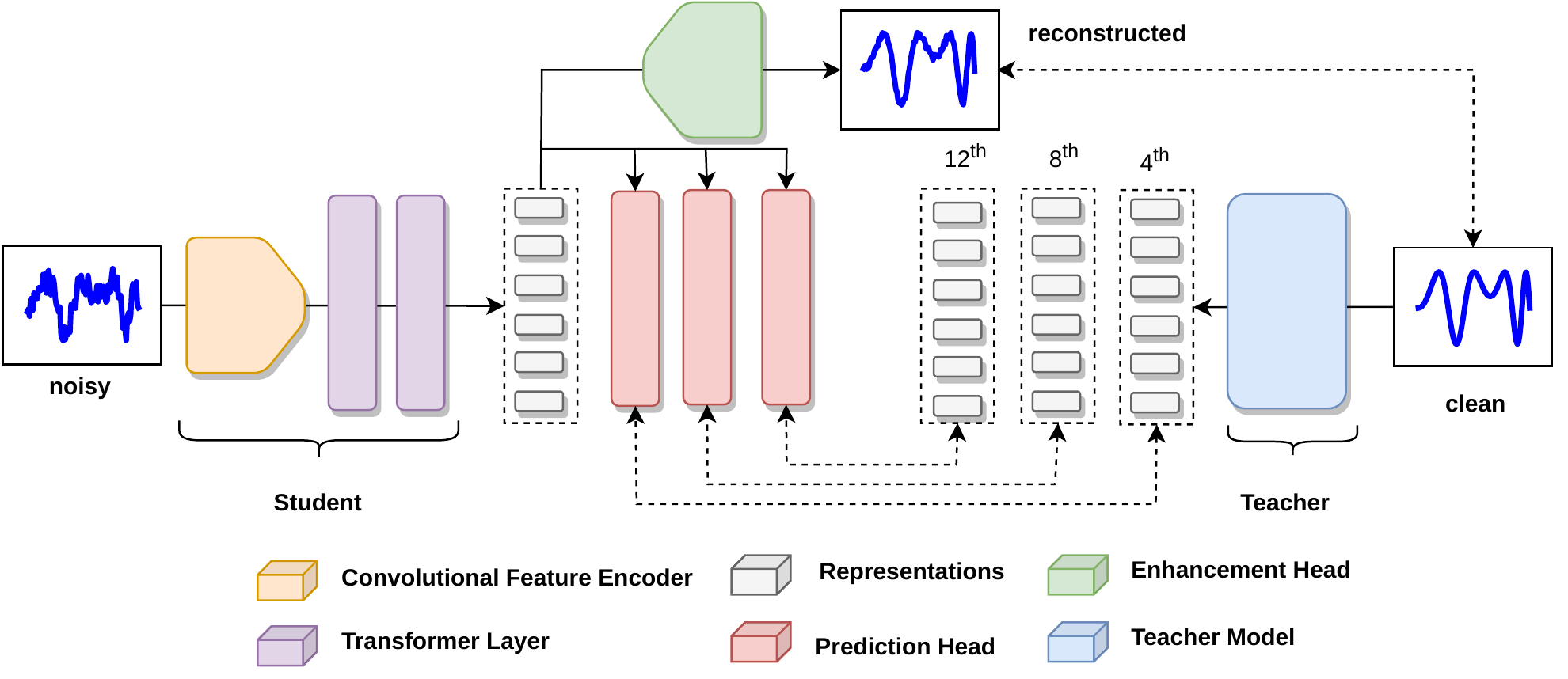} 
        \caption{Diagram of the proposed RobustDistiller recipe. 
        (a) is the noisy; (b) is the clean; and (c) the reconstructed signal.}
        \label{fig:robust_distiller_diagram} \vspace{-2mm}
\end{figure}
\vspace{-2mm}
\subsection{Modification \#1: Data Augmentation}
Earlier SRL models commonly relied on learning features from large unlabeled audiobook data, such as LibriSpeech \cite{panayotov2015librispeech} or LibriLight \cite{kahn2020libri}. However, even though these models can learn fundamental characteristics from speech signals, real-world deployment data often involve diverse channel conditions and environmental noises that harm system performance, a problem known as domain shift. To tackle this issue, data augmentation has been proposed as a technique to improve data diversity and reduce model bias \cite{huang22b_interspeech, chen2022wavlm, hsu2021robust}. Here, we propose to perform an online contamination of the data during the distillation process. In particular, the student model receives the noisy data as input, but the network's target is to reconstruct the clean representations of the teacher model. At training time, given a batch of clean speech utterances, we uniformly sample one action to be applied to each utterance in the batch: (i) no changes are made to the training utterance; (ii) contaminate the utterance with additive noise with signal-to-noise ratio randomly chosen from $[0, 20]$ dB; (iii) convolve the speech waveform with a random room impulse response; and (iv) apply both noise and reverberation at the same time.
\vspace{-2mm}
\subsection{Modification \#2: Multi-Task Denoising Learning}
As mentioned previously, speech enhancement is a standard processing method to remove noise from a speech signal, usually leading to perceptual quality improvements. Motivated by \cite{wang2022improving}, we propose a multi-task learning approach, where beyond learning to reconstruct the teacher's representations, we propose an additional enhancement head responsible for rebuilding the clean speech waveform from the learned representation. Figure \ref{fig:robust_distiller_diagram} shows the schematics for the proposed model with the additional enhancement head. In contrast to usual enhancement techniques, our objective is to enforce the upstream model to carry enough information about the speech itself and not the noise components, rather than optimize a quality metric. In this work, we design an enhancement head similar to the encoder of the student model. Here, a BiLSTM layer followed by seven transposed convolutions and GELU activation functions are used, as well as an L1-reconstruction loss between the enhanced and the clean waveform.

Both of these procedures together comprise what we call the RobustDistiller recipe which can be applied to any universal speech representation. For the experiments described herein, the recipe will be applied to distil the popular Wav2vec 2.0 \cite{baevski2020wav2vec}, HuBERT \cite{hsu2021hubert}, and WavLM \cite{chen2022wavlm} models.

\vspace{-3mm}
\section{Experimental Setup}
\label{sec:typestyle}
\vspace{-2mm}
\subsection{Datasets}
The dataset of clean speech utterances used to train the model is the LibriSpeech corpus  \cite{panayotov2015librispeech}, a dataset with 960 hours of audiobook recordings with a 16 kHz sampling rate derived from the LibriVox project. Additionally, to augment the training data, two noise datasets are used, namely MUSAN \cite{musan2015} and UrbanSound8K \cite{Salamon:UrbanSound:ACMMM:14}, as well as the OpenSLR28 dataset \cite{ko2017study} of room impulse responses (RIR). The noise datasets contain approximately 15 hours of recordings in a wide variety of categories, from office-like noises, natural sounds, babble noise, and urban sounds, such as engine idling and siren. We removed the children playing and street music categories to focus on non-speech like noise sources in this first analysis. All the utterances are resampled to 16 kHz. Lastly, the OpenSLR28 dataset contains 325 RIRs recorded in real environments, from small meeting rooms to large auditoriums.

At test time, multiple noise datasets are used to test the model performance under unseen conditions. Three separate datasets are used to this end. The first is the Acoustic Scene Classification from the Detection and Classification of Acoustic Scenes and Events (DCASE) Challenge of 2020, namely DCASE2020 \cite{Mesaros2018_DCASE}. The dataset consists of 64 hours of audio recordings in 10 acoustic scenes, recorded with four different recording devices in 12 different cities. For reverberation, we used the OpenSLR26 dataset \cite{ko2017study}, which contains 60,000 simulated room impulse responses corresponding to various small-, medium-, and large-sized rooms. Finally, we also rely on the Deep Noise Suppression Challenge 4 (DNS4) noise dataset \cite{dubey2022icassp} to measure the effectiveness of using speech enhancement models as a pre-processing tool. This dataset comprises 180 hours of noise spread across 62,000 utterances covering 150 different noise types. Here, it is important to emphasize that this dataset is not used when testing with the WavLM model to avoid data leakage.
\vspace{-2mm}
\subsection{Pre-training}
To gauge the benefits of our proposed methodology, different experiments are tested. First, we selected the wav2vec 2.0 base, HuBERT base, and WavLM base+ as our Teacher models for the distillation process. First, we apply the distillation process described in \cite{chang2022distilhubert}, resulting in the models we refer to as `Distilwav2vec 2.0', `DistilHuBERT', and `DistilWavLM', respectively. It is important to emphasize that we distilled the \{4\textsuperscript{th}, 8\textsuperscript{th}, and 12\textsuperscript{th}\} layers of all teacher models, which may not be ideal for models other than HuBERT \cite{chang2022distilhubert}. As such, the reported results may be a lower bound on the possible achievable accuracy. Furthermore, in the first set of (ablation) experiments, we explore the impact of only adding data augmentation to the distillation process. Lastly, we experiment with the multi-task learning paradigm by attaching the speech enhancement head to the pre-training step.

In all experiments, the upstream models are trained using a single NVidia A100 GPU. Our method takes approximately 30 hours to train when using data augmentation, whereas experiments using the denoising head take roughly 43 hours. We use the AdamW optimizer, with a batch of 24 utterances, for 200k iterations, whereas after 14k updates, the learning rate linearly decays from $2\times10^{-4}$ to zero.

\begin{table*}
    \centering
    \caption{Experimental results under clean (\textbf{c}), noisy (\textbf{n}), reverberation (\textbf{r}), and noise-plus-reveberation (\textbf{n+r}) test conditions.}
    \label{tab:results}
    \resizebox{\textwidth}{!}{%
    \begin{tabular}{cccccccccccccccc}
        \toprule
        & & & \multicolumn{4}{c}{Keyword Spotting (KS)} & \multicolumn{4}{c}{Intent Classification (IC)} & \multicolumn{4}{c}{Emotion Recognition (ER)} & \multirow{2}{*}{\textbf{Overall}} \\

        \cmidrule(lr){4-7}
        \cmidrule(lr){8-11}
        \cmidrule(lr){12-15}

        & Upstream & \#params (M) & (c) & (n) & (r) & (n + r) & (c) & (n) & (r) & (n + r) & (c) & (n) & (r) & (n + r) &   \\ 

        \midrule
        \multirow{5}{*}{\rotatebox[origin=c]{90}{\large Baselines}} & wav2vec2 Base & 95 & 96.20 & 86.50 & 64.43 & 56.25 & 92.83 & 69.26 & 68.78 & 50.88 & 63.01 & 51.46 & 32.70 & 27.02 & 63.28 \\
        & HuBERT Base & 95 & 96.30 & 84.55 & 61.70 & 51.61 & 98.34 & 78.70 & 75.77 & 59.08 & 64.75 & 53.45 & 40.72 & 29.12 & 66.17 \\
        & WavLM Base+ & 95 & 96.88 & 90.07 & 78.64 & 72.38 & \textbf{98.84} & 83.13 & \textbf{89.35} & 68.20 & \textbf{67.96} & 55.08 & 36.57 & 27.14 & 72.02 \\
        & Robust HuBERT & 95 & 96.33 & 92.15 & 74.81 & 66.50 & 98.66 & \textbf{91.43} & 85.32 & 71.10 & 64.59 & 57.17 & \textbf{41.44} & 33.36 & 72.74 \\
        & LightHuBERT & 27 & 96.11 & 86.69 & 72.35 & 62.25 & 98.60 & 83.73 & 88.19 & 70.45 & 63.88 & 48.72 & 35.51 & 29.84 & 69.69 \\[1mm]
        \midrule
        
        \multirow{3}{*}{\rotatebox[origin=c]{90}{\large Distiller}} & Distilwav2vec2 & 24 & 95.33 & 82.21 & 55.18 & 45.37 & 86.05 & 39.81 & 29.24 & 17.37 & 61.01 & 49.94 & 24.14 & 23.69 & 50.78 \\
        & DistilHuBERT & 24 & 96.14 & 86.34 & 59.20 & 54.53 & 94.12 & 48.72 & 47.72 & 24.31 & 62.55 & 47.89 & 31.35 & 25.77 & 56.55 \\
        & DistilWavLM & 24 & 96.36 & 88.48 & 61.34 & 57.58 & 94.89 & 63.25 & 48.43 & 31.69 & 62.09 & 51.14 & 29.24 & 25.27 & 59.15 \\[1mm]
        
        \midrule

        \multirow{6}{*}{\rotatebox[origin=c]{90}{\large RobustDistiller}} & \textbf{Robust Distilwav2vec2} & 24 & 95.75 & 91.46 & 86.01 & 79.81 & 89.74 & 70.23 & 73.32 & 56.84 & 62.44 & 50.53 & 41.18 & 30.73 & 69.00 \\
        & \qquad - w/ enhancement head & 24 & 96.20 & 92.37 & 84.88 & 79.03 & 89.11 & 75.32 & 73.79 & 61.11 & 62.37 & \textbf{56.28} & 40.12 & 32.75 & 70.28 \\
        & \textbf{Robust DistilHuBERT} & 24 & 96.59 & 93.02 & 85.98 & 80.33 & 94.15 & 82.52 & 83.42 & 65.38 & 61.71 & 49.98 & 40.83 & 30.00 & 71.99 \\
        & \qquad - w/ enhancement head & 24 & 96.69 & 93.12 & 85.98 & 80.04 & 94.57 & 83.89 & 82.75 & 70.29 & 61.88 & 53.59 & 39.27 & 28.74 & 72.57 \\
        & \textbf{Robust DistilWavLM} & 24 & 96.46 & 93.44 & \textbf{88.70} & \textbf{83.09} & 94.81 & 81.60 & 87.21 & \textbf{72.63} & 61.66 & 51.80 & 39.66 & \textbf{33.26} & \textbf{73.69} \\
        & \qquad - w/ enhancement head & 24 & \textbf{96.98} & \textbf{93.87} & 86.21 & 82.15 & 95.07 & 80.33 & 84.16 & 71.18 & 62.99 & 55.90 & 37.15 & 29.29 & 72.94 \\[1mm]
        \midrule[\heavyrulewidth]
        \bottomrule
    \end{tabular} \vspace{-6mm}
    }
\end{table*}

\vspace{-2mm}
\subsection{Downstream Tasks and Figure-of-Merit}
Here, we use a subset of three downstream tasks from the SUPERB benchmark \cite{yang2021superb} to evaluate the robustness of the proposed modifications: keyword spotting, intent classification, and emotion recognition. For all downstream tasks, the evaluation metric is the standard accuracy score; thus, higher values are better. Four evaluation scenarios are considered in our analyses: {clean} $(c)$, {noise-only} $(n)$, {reverberation-only} $(r)$, and {noise-plus-reverberation} $(n+r)$. The test set meets the same criteria as SUPERB downstream tasks in the clean setting. Additive noise with signal-to-noise ratios ranging from $[-5, 20]$ dB is added for the noisy test conditions. For the reverberation condition, a random room impulse response is uniformly sampled and applied to the test set utterance. Lastly, for the noise-plus-reverberation condition, both noise and reverberation are jointly applied to the test set, thus representing the most difficult scenario. A custom seed is applied to ensure all models are evaluated with the same degradations.

\vspace{-2mm}
\section{Experimental Results and Discussion}
\label{sec:majhead}
\vspace{-2mm}
\subsection{Classification accuracy}
Table~\ref{tab:results} presents the results achieved with i) the three baseline teacher models, along with a robust version of HuBERT and its `Light' version, compressed to 27 million parameters \cite{wang22t_interspeech}, ii) the distilled models following the conventional distillation process proposed in \cite{chang2022distilhubert}, and iii) the proposed distillation methodology. Results are reported across the four testing scenarios (clean, noise-only, reverberation-only, and noise-plus-reverb) for three downstream tasks. As can be seen, the proposed RobustDistiller methodology outperforms the original Distiller recipe for the three SUPERB tasks across all scenarios, even in the clean condition. Notably, the RobustDistiller paradigm applied to the WavLM Base+ teacher model achieved the best overall accuracy among all models. In fact, on average it even outperformed the original teacher model with roughly four times as many parameters. In fact, just adding data augmentation improved clean speech accuracy for the KS and IC tasks for the robust distilled models from wav2vec2.0 and HuBERT, thus showing the benefits of diverse data on the distillation process.

Coupled with data augmentation, the additional enhancement head proved to be especially useful in clean and noisy settings, further improving the results of RobustDistiller. Furthermore, we hypothesize that the multi-task learning paradigm helps the model to better generalize by reinforcing the disentanglement of speech features from the noise factors, thus achieving better reconstructions for the clean waveform given the compressed representation. Overall, for KS the RobustDistiller with only 24M parameters has results inline with or outperforming those of the larger teacher models across testing conditions. For the noise-plus-reverberation scenario, in turn, the proposed  RobustDistilWavLm model  achieved results inline with the robust larger models (WavLM and Robust HuBERT) for the other two tasks.




One question that may arise is the need to create robust models if state-of-the-art speech enhancement algorithms exist and could be used as a pre-processing step prior to the upstream task. To show the advantages achieved with the proposed method, Table~2 depicts the accuracy achieved for the keyword spotting task once the DNS4 dataset was used to corrupt the test set. Two denoising algorithms are explored, namely MetricGAN+~\cite{fu21_interspeech} and SepFormer~\cite{sepformer}. The former is an enhancement method that optimizes a non-differentiable quality metric (i.e., the Perceptual Evaluation of Speech Quality, PESQ) through a metric-learning approach in an adversarial training fashion, where the discriminator that learns to mimic the behavior of the PESQ evaluation metric through a neural network. 
The latter, in turn, is a Transformer-based neural network for source separation that receives as input a raw waveform and learns local and global structures from audio. 
The SepFormer is trained on the WHAMR!~\cite{maciejewski2020whamr} corpus, which contains noise and reverberation. As can be seen, enhancement at the signal level can degrade classification accuracy relative to processing the noisy signal directly. This is likely due to unwanted artifacts that can act as distribution shifts in the data, or to the fact that enhancement may remove soome of the discriminatory segments of the data used by the representation. Comparisons between the results in Tables 1 and 2 show the advantages of the proposed method over conventional speech enhancement. 

\begin{table}
\small
    \centering
    \label{tab:results_enhanc}
    \caption{Experimental results with speech enhancement.}
    \vspace{0.1in}
    \begin{tabular}{ccccc}
        \toprule
        \multirow{2}{*}{Upstream}& \multicolumn{3}{c}{Keyword Spotting (KS)} \\
        \cmidrule(lr){2-4}
         & Noisy & MetricGAN+ & SepFormer \\ 
        \midrule
        wav2vec2 Base  & 83.84 & 71.31 & 78.61 \\ 
        HuBERT Base & 84.10 & 74.13 & 80.85 \\   
        DistilHuBERT & 83.45 & 72.44 & 81.17 \\ 
        LightHuBERT & 82.80 & 73.09 & 80.30 \\ 

        \midrule[\heavyrulewidth]
        \bottomrule
    \end{tabular} 
\end{table}

\vspace{-2mm}
\subsection{Effects of Noise Type}
\begin{figure}
   \centering
   \includegraphics[width=0.91\linewidth]{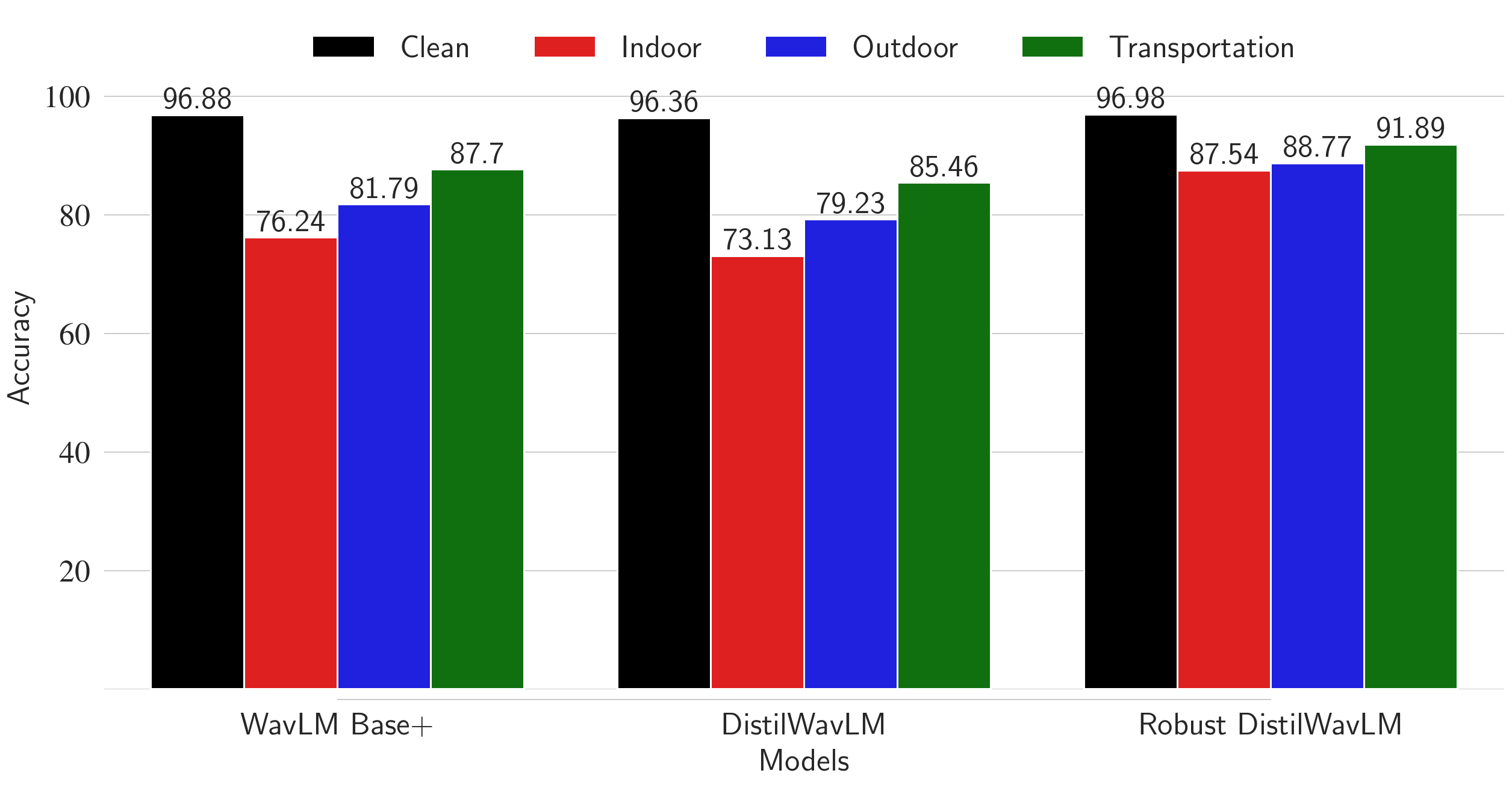}
   \caption{KS accuracy per noise types. Clean bars for reference.}
   \label{fig:noise_env} \vspace{-3mm}
\end{figure}

From Table 1, the detrimental effect of noise can be seen on system performance; careful analysis has shown that this was particularly true at low SNR regimes below an SNR of 10 dB. As such, here we are interested in gauging if noise type also plays a role on final performance. To this end, 
we utilize the DCASE2020 challenge dataset comprised of noise types representative of three categories: indoor (e.g., shopping mall, airport), outdoor (e.g., park, public square), and transportation (e.g., bus, metro, tram). In particular, focus is placed on the KS task and noise is added to the test set at SNR in $[-5,5]$ dB range. Figure~\ref{fig:noise_env} depicts the accuracy achieved for the WavLM Base+ teacher model, its conventional distilled version, as well as the proposed version. As can be seen, indoor noises provide the largest degradation, likely due to the combined reverberation and additive noise effects, followed by outdoor and transportation, respectively. Overall, the proposed recipe provides not only the largest accuracy across all categories, but also the most stable results.

\vspace{-3mm}
\subsection{Effect of Room Reverberation}
\begin{figure}
   \centering
   \includegraphics[width=0.7\linewidth]{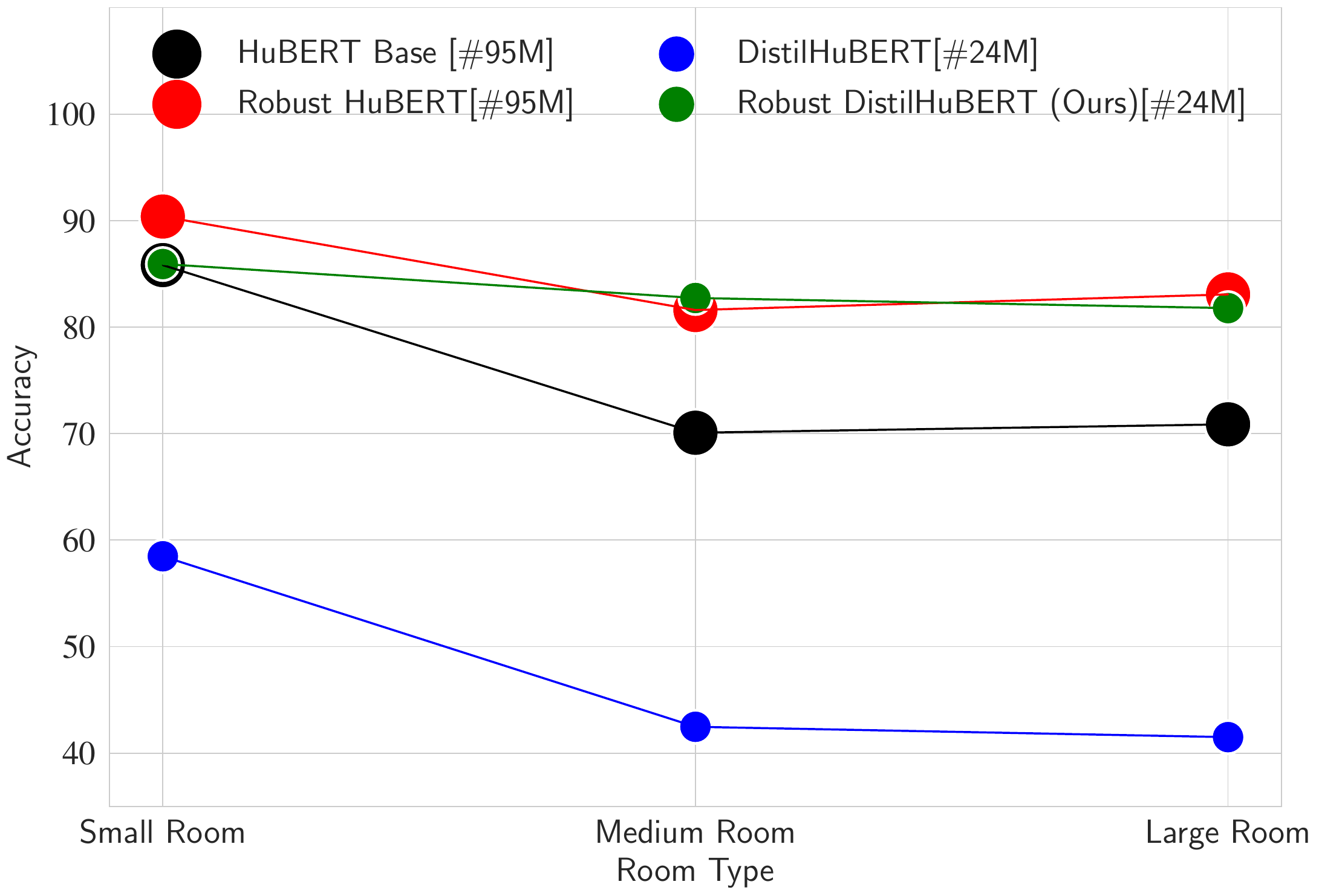}
   \caption{IC performance per room size.}
   \label{fig:reveb_env} \vspace{-3mm}
\end{figure}

\vspace{-1mm}
Lastly, we are interested in further gauging the impact of different room sizes (i.e., larger reverberation times) on system performance; here, the intent classification task is used. Figure~\ref{fig:reveb_env} summarizes the obtained results for test samples corrupted by RIRs representative of small-, medium- and large-sized rooms. As can be seen, medium and large sized rooms resulted in accuracies lower than those achieved with smaller rooms. Overall, the proposed Robust DistilHuBERT model with 24M parameters achieved the best accuracies and inline with those obtained with the larger (95M parameter) published Robust HuBERT model. Such findings are very promising and suggest that robust distillation could indeed be achieved for edge speech applications.

\vspace{-3mm}
\section{Conclusions}
\label{sec:print}
\vspace{-2mm}
Here, we propose RobustDistiller, a methodology to improve the layer-wise distillation process to provide a more robust universal speech representations. In particular, two modifications are applied to the DistilHuBERT recipe: addition of data augmentation and multi-task learning, where a speech enhancement step is jointly performed. Experiments on three SUPERB tasks show the proposed model outperforming the original distiller recipe across clean and noisy conditions, oftentimes outperforming the original teacher models. Overall, the proposed RobustDistiller recipe on top of the WavLM Base+ teacher model resulted in the best model both in terms of accuracy and robustness, suggesting this could be a suitable candidate for future edge speech applications.



\bibliographystyle{IEEEbib}
\bibliography{strings,refs}

\end{document}